\documentclass[letterpaper]{article}
\usepackage{tabularx} 
\usepackage{amsmath}  
\usepackage{graphicx} 
\usepackage[margin=1in,letterpaper]{geometry} 
\usepackage{cite} 
\usepackage[final]{hyperref} 
\hypersetup{
	colorlinks=true,       
	linkcolor=blue,        
	citecolor=blue,        
	filecolor=magenta,     
	urlcolor=blue         
}

\begin{document}

\title{An Extension of Deep Pathway Analysis: A Pathway Route Analysis Framework Incorporating Multi-dimensional Cancer Genomics Data}
\author{Yue Zhao}
\maketitle

\abstract{\textbf{Motivation:} Recent breakthroughs in cancer research have come via the up-and-coming field of pathway analysis. By applying statistical methods to prior known gene and protein regulatory information, pathway analysis provides a meaningful way to interpret genomic data. While many gene/protein regulatory relationships have been studied, never before has such a significant amount data been made available in organized forms of gene/protein regulatory networks and pathways. However, pathway analysis research is still in its infancy, especially when applying it to solve practical problems. \\
	\textbf{Results:} In this paper we propose a new method of studying biological pathways, one that cross analyzes mutation information, transcriptome and proteomics data.  Using this outcome, we identify routes of aberrant pathways potentially responsible for the etiology of disease. Each pathway route is encoded as a bayesian network which is initialized with a sequence of conditional probabilities specifically designed to encode directionality of regulatory relationships encoded in the pathways. Far more complex interactions, such as phosphorylation and methylation, among others, in the pathways can be modeled using this approach. The effectiveness of our model is demonstrated through its ability to distinguish real pathways from decoys on TCGA mRNA-seq, mutation, Copy Number Variation and phosphorylation data for both Breast cancer and Ovarian cancer study. The majority of pathways distinguished can be confirmed by biological literature. Moreover, the proportion of correctly indentified pathways is \% higher than previous work where only mRNA-seq mutation data is incorporated for breast cancer patients. Consequently, such an in-depth pathway analysis incorporating more diverse data can give rise to the accuracy of perturbed pathway detection.\\
	\textbf{Contact:} \href{yue.2.zhao@uconn.edu}{yue.2.zhao@uconn.edu}\\

\maketitle

\section{Introduction}
Pathway analysis has been playing an important role in cancer research. By and large, pathway analysis helps interpret genomics data by applying computational methods which combine prior known gene/protein regulatory information. There exists ample gene/protein regulatory relationships summarized in the literature that is organized into various forms of gene/protein regulatory networks and pathways. However, pathway analysis research is still a nascent area, particularly when it comes to practical problems. For instance, breast cancer patients with the ERBB2 amplification are normally cured by Trastuzumab. According to \cite{vogel2001first}, less than fifty percent of the patients get the benefit of the ERBB2 targeting antibody, indicating a deeper understanding of a patient\rq s   pathway behavior is required. Therefore, developing a more comprehensive way to analyze pathways by combining multiple genomic data sets,  which are now readily available through various high-throughput sequencing technologies (e.g., RNA-Seq, DNA-Seq, ChIP-Seq), is of great signficance. \cite{zhao2016deep} propose an approach by modeling the pathway route as an analysis unit.  Nevertheless, this approach has many defects:
\begin{itemize}
	\item missing values are not penalized
	\item more data types like Copy Number Variation (CNV) and DNA methylation data are not included in the model
	\item specific types of regulation information are not considered, such as phosphorylation, methylation
\end{itemize} 
The goal of this paper is to extend that pathway analysis framework giving it the ability to include proteomics and CNV data and the specific types of regulation mentioned above.  Together with existing transcriptome and mutation data, we aim to pinpoint the precise pathway routes perturbed.
This analysis will focus on explaining the biological mechanisms behind cancer development more accurately. The rest of the paper is outlined as follows. In section II, existing pathway analysis methods are briefly reviewed. Section III covers the model settings and assumptions described in detail. Section IV presents a significance study similar to that of \cite{vaske2010inference} using TCGA Breast Cancer data. In section V, we show the route analysis outcome of applying our model to the data. Finally we conclude with section VI.

\section{Related Work}
Great efforts have been made to incorporate pathway information into genomic data analysis. One of the first popular methods of analyzing genome-wide experimental data is using gene set enrichment analysis methods \cite{subramanian2005gene}. \cite{li2008network} encoded the pathway network into a penalty function and did model selection by optimizing the function to pick meaningful genes and subnetworks.  \cite{tarca2009novel} proposed SPIA which measures pathway significance by statistical testing against random permutation. \cite{vaske2010inference} presented PARADIGM, a novel method by modeling the pathway as a factor graph to do patient specific inference.  \cite{verbeke2015pathway} ranked the pathways by p-value obtained from encoding pathway logic into a global network. The p-value is calculated based on a hypothesis test where the null hypothesis is that the pathway is picked randomly. \cite{korucuoglu2014bayesian} and \cite{isci2011pathway} encode the pathway as a Bayesian network. After removing cycles in the graph, they train the model with expression data. Significance of the score is given by bootstrap-generated data. \cite{zhao2016deep} dynamically encode pathway routes as a Bayesian network incorporating expression and mutation data to do the pathway analysis.

	\section{Methods}
	
	\subsection{Model and Terms}
	
	
	\begin{figure}[!t]
		\centering
		\includegraphics[trim=0cm 0cm 0cm 0cm,clip,width=0.5\textwidth]{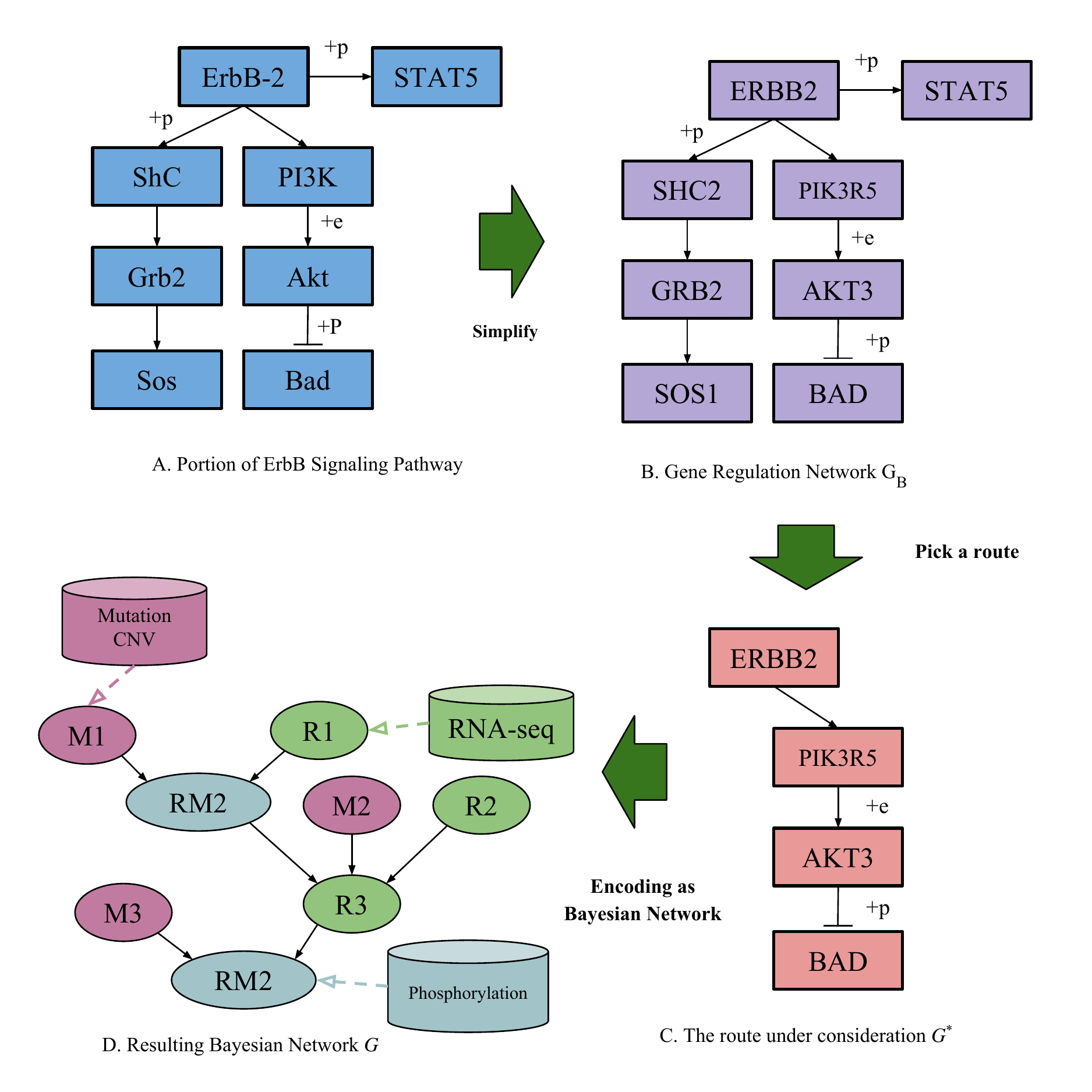}
		\caption{Conversion Pipeline. Part of the ErbB pathway in KEGG shown in part A. The pathway is simplified by keeping only specific interactions and genes, resulting in gene regulation network $G_B$ in Part B. A route $G^*$ shown in part C, starting from ERBB2 to BAD, is extracted from $G_B$ and converted to a Bayesian Network $G$ in part D. Since ERBB2 is Activating (one of Functional Interactions) PIK3R5 in $G^*$, thus the nodes for ERBB2, $R_1$ and $M_1$ is pointing to $RM_2$, the $RM$ node for PIK3R5. On the other hand, once getting activated, PIK3R5\rq s protein will binds to AKT3 and make it express. Thus all nodes for PIK3R5 points to $R$ node of AKT3, namely, $R_3$. Distinct node colors represent different data source (denoted as cylinders) used to calculate the observations. For instance, all $R$ node observation comes from mRNA-seq database for the same patient.}
		\label{p2_pipeline}
	\end{figure}
	
	Figure~\ref{p2_pipeline} illustrates the pipeline of this approach. Figure~\ref{p2_pipeline}.A shows an example pathway, ErbB, which has been adapted from the KEGG pathway database \cite{kanehisa2000kegg}. The pathway is then simplified to be a gene regulation network $G_B$.  Unlike most existing approaches who merely keep activation and inhibition interactions after the simplification, different tags on the interactions in the KEGG pathway are furthermore kept in $G_B$, i.e. Phosphorylation (+p), Ubiquitination(+u), Glycosylation(+g), Methylation(+m), Dephosphorylation(-p), Debiquitination(-u),  Deglycosylation(-g), Demethylation(-m), Expression(+e), Repression(-e). These tags are defined as evidence tags determining the data associated with the interactions.

	Among all these types of tags, Phosphorylation, Glycosylation, Methylation, Ubiquitination, Dephosphorylation, Deubiquitination,  Deglycosylation, Demethylation do not affect the expression level of the target gene, however, it can affect the protein product structure and function. We call all these types of edges (Protein) \textbf{Functional Interactions}. For Expression and Repression edges, only expression level of $g_i$ is affected rather than the protein function. Thus, we categorize these two type of interactions as \textbf{Expression Interactions}. These two categories will be handled in distinct ways in the model.
	
	The next step is to identify all possible ``routes'' available from the given $G^*$. Figure~\ref{p2_pipeline}.C shows a route which starts from ERBB2 and ends at BAD. The selected route is then converted into a discrete Bayesian Network ( denoted as $G$ shown by Figure~\ref{p2_pipeline}.D). Our objective is to treat each ``route'' as a unit of pathway analysis. This route-based method is more amenable as it could show whether the effect of ERBB2 amplification is more prominent through ERBB2$\rightarrow$SHC2 path or ERBB2$\rightarrow$PIK3R5 path, or even for both. The route-based modeling idea assumes that it is crucial to identify which portion(s) of the pathway is(are) either abnormally activated or suppressed. In this way, a more informed treatment plan could be designed.
	
	We then continue to illustrate the conversion process from a pathway route $G^*$ to a Bayesian network $G$. As shonw in Fig.\ref{p2_bnEncoding}, for a gene regulation network $G_B$ (converted from a pathway), a path $G^*$ is simply a subgraph of $G_B$, $G^*\subseteq G_B$, $G^*=(V^*,E^*)$ where $V^*=\{g_1,\ldots,g_{k_{G^*}}\}$, where $g_i$ represents the $i$th gene and $k_{G^* }$ is the number of genes contained in path $G^*$, $E^*=\{e_{ij} |1\leq i<k_{G^* }$ and $j=i+1\}$. 
	
	For each edge in $G^*$, $e_{i-1,i},1<i\leq k_{G^*}$, if $i<k_{G^*}$ and $e_{i-1,i}$ is one of the Functional Interactions in $G^*$, then we create three nodes in the corresponding Bayesian Network $G:$ $R_i$, $M_i$ and $RM_i$ for $g_i$. On the other hand, if $i<k_{G^*}$ and $e_{i-1,i}$ is one of the Expression Interactions, only two nodes $R_i$ and $M_i$  will be created. The first gene, $g_1$  will always have two nodes created $R_1$ and $M_1$ while $g_{k_{G^*}}$ will only have one node, either $R_{k_{G^*}}$ (if $e_{k_{G^*}-1,k_{G^*}}$ is Expression Interaction) or $RM_{k_{G^*}}$  (if $e_{k_{G^*}-1,k_{G^*}}$ is Functional Interaction). In this way, there will usually be three nodes for target gene of Functional interactions and two nodes for that of Expression interactions.
	
	After creating nodes for each gene in the path $G^*$, the edges in the Bayesian network $G$ will be added dynamically according to the edges in pathway route $G^*$.  For $g_i \in V^*, 1<i\leq k_{G^*}$,  if $e_{i-1,i}\in E^*$ is one of the Functional Interactions, edges will be created pointing from all the nodes for parent gene $g_{i-1}$ to the $RM$ node child gene $g_i$. Namely, we will add  edges from $R_{i-1},M_{i-1},(RM_{i-1})$, to $RM_i$. On the other hand if $e_{i-1,i}$ is an Expression Interaction, edges from the nodes of $g_{i-1}$ ($R_{i-1},M_{i-1},(RM_{i-1})$) to $R_{i}$ will be created instead.
	The conditional probability table corresponding to edges in Bayesian Network $G$ is determined by the type of the edge in $G^*$ as shown in details from Table~\ref{expression_cpt} to Table~\ref{functional_minus_cpt}. The assumption is that, given the edge $e_{i-1,i}$, the expression level ($R_i$) (or functional status $RM_i$), of the gene $g_i$ is affected by its parent\rq s expression status $R_{i-1}$, the DNA functional status $M_{i-1}$ (and the Protein functional status $RM_{i-1}$ if exists). After conversion, the resulting Bayesian Network $G$ is formally defined as follows: $G=(V,E)$, where $V=RR\bigcup MM \bigcup RMS$, $RR=\{R_i, i\in \{1,\ldots,k_{G^*}\}\}$ where $R_i$ is a random variable representing expression level status on gene $g_i$. $MM=\{M_i, i\in \{1,\ldots,k_{G^*}-1\}\}$ where  $M_i$ is a random variable representing DNA functional status on gene $g_i$. $RMS=\{RM_i,i\in \{j:e_{j-1,j}\text{ is one of Functional Interactions in }G^*\}\}$ where $RM_i$ is a random variable representing protein functional status on gene $g_i$. This setting is motivated from central dogma shown in Figure~\ref{central_dogma}.
	
	\begin{figure}[!t]
		\centering
		\includegraphics[trim=0cm 1.3cm 0cm 0.3cm,clip,width=0.5\textwidth]{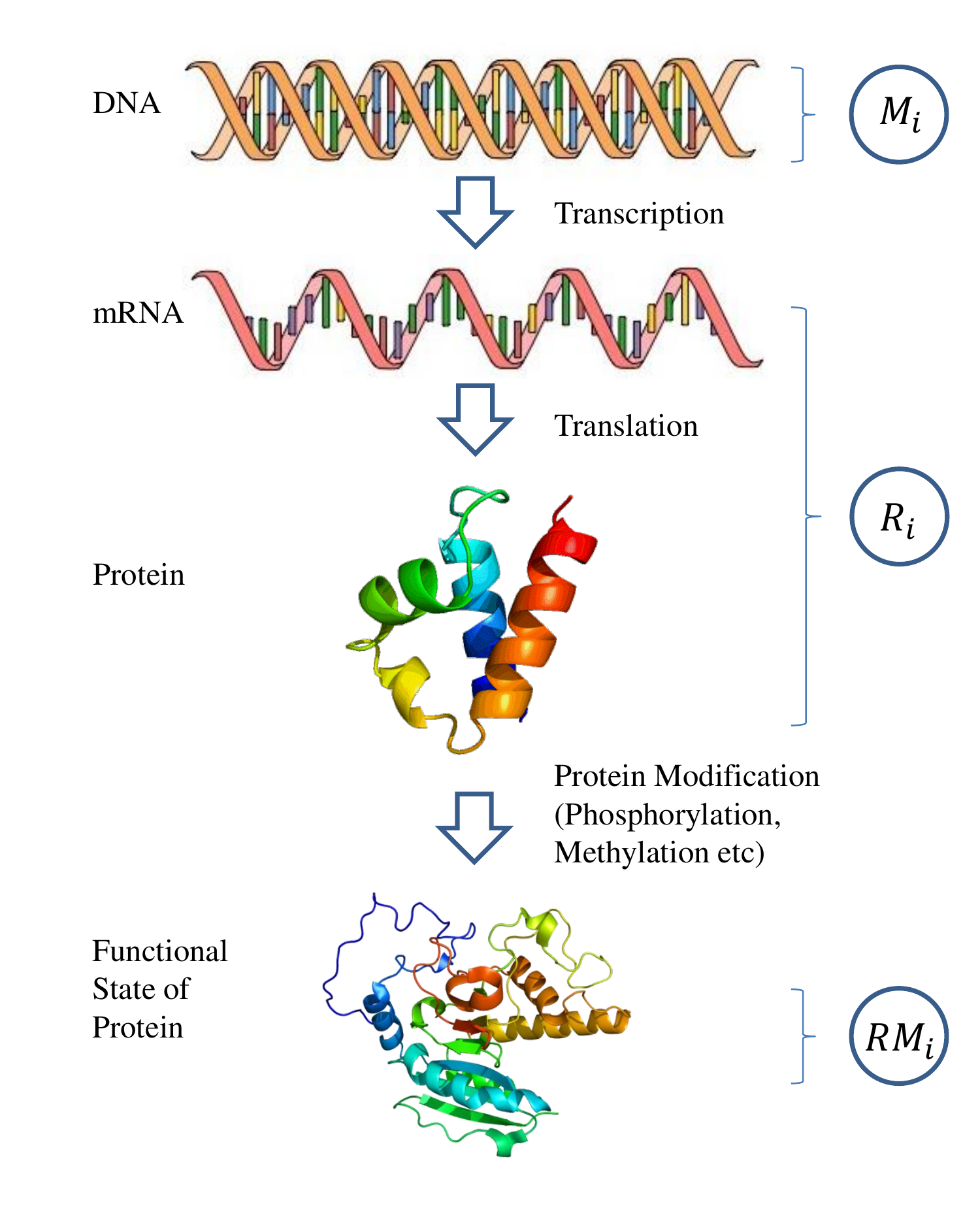}
		\caption{Central Dogma of Molecular Biology for gene $g_i$. Random variable $M_i$s will be handling the DNA functional status. Mutation and Copy Number Variation data will be used to get the observation for the $M$ random variable. Random variable $R_i$ will be representing the expression status, namely, whether the mRNA or protein of $g_i$ will be high enough be achieve the biological function. The last type of random variable $RM_i$ stands for the functional status of the protein, for example, whether the protein has been phosphorylated or methylated by $g_i$\rq s parent gene. Only when all three parts work properly, could the interaction $e_{i+1,i+2}$ take effect. }
		\label{central_dogma}
	\end{figure}
	
	\begin{figure}[!t]
		\centering
		\includegraphics[trim=0cm 0cm 0cm 0cm,clip,width=0.5\textwidth]{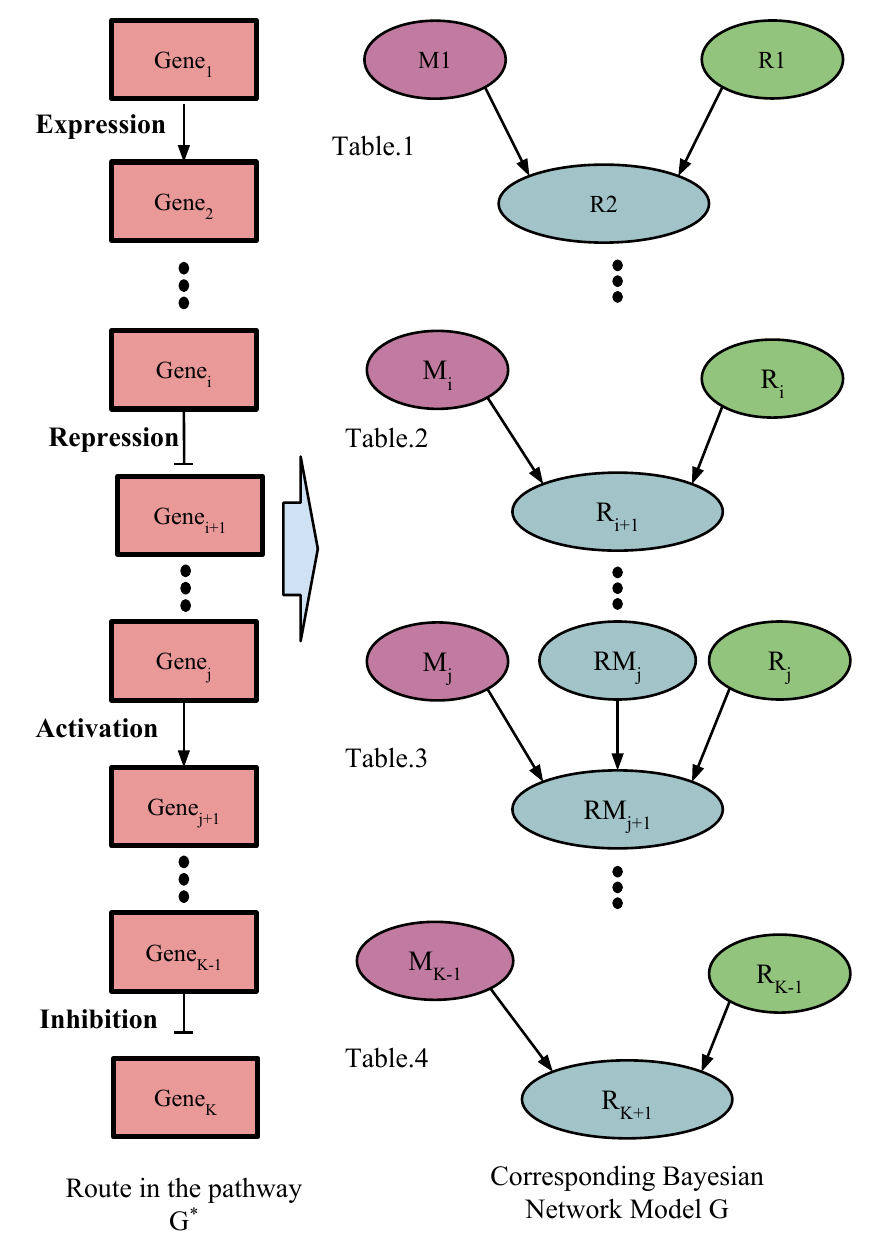}
		\caption{Converting the route in the pathway to Bayesian Network. The pathway route $G^*$ on the left is converted to Bayesian Network $G$ on the right. Especially, for $Gene_j$ activating $Gene_{j+1}$, the $RM_{j+1}$ has an extra parent $RM_{j}$. This illustrates the special case when $Gene_{j-1}$ is activating or inhibiting $Gene_j$. }
		\label{p2_bnEncoding}
	\end{figure}

	
	$M_i$, $R_i$ and $RM_i$ are now defined in detail. Since the DNA information is not affected by any interactions in the pathway route and $M_i$ doesn\rq t have a parent node in $G$, the random variable $M_i$ follows a Bernoulli distribution as shown in (\ref{mut_prior}). The Bernouli random variable $M_i$ has two possible values: $+1$ represents that $g_i$ functions normally on DNA level, and $-1$ represents function loss, i.e. $g_i$\rq s DNA original biological function is disrupted. The probability distribution indicates the prior has no specific preference on these two levels:
	\begin{equation}
	\label{mut_prior}
	M_i=\begin{cases}
	+1 & p=0.5 \\
	-1 & p=0.5
	\end{cases}
	\end{equation}
	Random variable $R_i$ follows a different probability distribution based on the location of gene  $g_i$ in path $G^*$: Suppose $g_i$ is the starting node in $G^*$, $R_i$'s distribution is shown in (\ref{rna_prior})
	\begin{equation}
	\label{rna_prior}
	R_i=\begin{cases}
	+1 & p=1/3 \\
	0 & p=1/3 \\
	-1 & p=1/3
	\end{cases}
	\end{equation}
	where +1 represents gene $g_i$ is not down regulated in expression level, $-1$ represents gene $g_i$ is down regulated in test cases and $R_i=0$ otherwise. For cancer research, test case is equivalent to the tumor cell. On the other hand, if $g_i$ ($i>1$) has a parent, gene $g_{i-1}$ in $G^*$, $R_i$ will follow the conditional probability table in  Table~\ref{expression_cpt}(Table~\ref{repression_cpt}) if $e_{i-1,i}$ is expression(repression) in $G^*$. In order to illustrate our model more clearly, we define the following operator $\&$, which is really similar to the AND operator, in (\ref{and_operator}). For instance, ($RM_{i-1}\&\text{)}M_{i-1}\&R_{i-1}$ has value of $+1$ if none of the three (or two if there is no $RM_{i-1}$) variables are $-1$. Otherwise, it has value of $-1$. 
	\begin{equation}
	\label{and_operator}
	A_1\& A_2 \&,\ldots,A_{n-1}\& A_{n}=\begin{cases}
	-1 & \exists i\in [1,n]\ s.t. \ A_i=-1 \\
	+1 & otherwise \\
	
	\end{cases}
	\end{equation}

	Next we show the biological logic behind the conditional probability table for $R_i$. Here we focus on the expression table (Table.\ref{expression_cpt}); the repression table (Table.\ref{repression_cpt}) is built in a similar way. If the parent gene of $g_i$, $g_{i-1}$, has no function loss in DNA, over-expresses and the functional status of $g_{i-1}$\rq s protein is fully activated, namely $M_{i-1}\&R_{i-1}\&RM_{i-1}=+1$, then the target $g_i$ will also be highly likely to overexpress, i.e. $R_i=+1$, given the edge between them in $G^*$ is \lq expression\rq. If there is no Functional Interaction $e_{i-2,i-1}$ targeting at $g_{i-1}$, there will be just $R_{i-1}$ and $M_{i-1}$ in the conditional table. As a result, $$P(R_i=+1|M_{i-1}\&R_{i-1}\&RM_{i-1}=+1)=1-\epsilon_1-\epsilon_2 $$ while $$P(R_i=-1|M_{i-1}\&R_{i-1}\&RM_{i-1}=+1)=\epsilon_1$$ $$P(R_i=0|M_{i-1}\&R_{i-1}\&RM_{i-1}=+1)=\epsilon_2$$ where $\epsilon_1$ and $\epsilon_2$ are respectively the probability of observing $R_i=-1$ and $R_i=0$. $1-\epsilon_1-\epsilon_2$ should be close to 1. Here $\epsilon_1<\epsilon_2$ indicating that we penalize the inconsistency more than the uncertainty. Similarly, if the parent gene of $g_i$ has DNA function loss, caused by mutation for instance, or its expression level is down regulated in test case, or the protein of $g_{i-1}$ is not activated successfully ($M_{i-1}\&R_{i-1}\&RM_{i-1}=-1$), then the downstream regulation process towards $g_i$ is likely not to be functioning. Therefore, $g_i$ would tend to be down regulated, namely $R_i=-1$, and hence the corresponding probability would be flipped.

	Similar to $R_i$, Random variable $RM_i$ has three possible values: $\{+1,0,-1\}$, where +1 represents gene $g_i$ has its protein switched on by its parent gene $g_{i-1}$ through $e_{i-1,i}$, $-1$ represents gene $g_i$ is switched off and otherwise $R_i=0$. Recall that $RM_i$ will be attached only when interaction $e_{i-1,i}$ in $G^*$ is Functional Interaction, $g_i$ will always have a parent, gene $g_{i-1}$ in $G^*$.  $RM_i$ will follow the conditional probability table from Table~\ref{functional_plus_cpt} to Table~\ref{functional_minus_cpt}. 
	
	The biological logic behind the conditional probability table for $RM_i$ is built based on central dogma, as shown in Figure~\ref{central_dogma}. Here we focus on the positive effect table (Table.\ref{functional_plus_cpt}); the negative effect table (Table.\ref{functional_minus_cpt}) is built in a similar way. If the parent gene of $g_i$, $g_{i-1}$, has no function loss in DNA, it over-expresses and $g_{i-1}$\rq s protein is successfully regulated (if $RM_{i-1}$ exists) ($M_{i-1}\&R_{i-1}\&RM_{i-1}=+1$), then interaction $e_{i-1,i}$ will take effect, thus the target $g_i$ protein will also be highly likely to be regulated successfully, namely, $R_i=+1$ given the edge between them in $G^*$ is one the Positive Functional Interactions: Phosphorylation, Glycosylation, Methylation, Ubiquitination, Activation . As a result, $$P(RM_i=+1|M_{i-1}\&R_{i-1}\&RM_{i-1}=+1)=1-\epsilon_1-\epsilon_2 $$ while $$P(RM_i=-1|M_{i-1}\&R_{i-1}\&RM_{i-1}=+1)=\epsilon_1$$ $$P(RM_i=0|M_{i-1}\&R_{i-1}\&RM_{i-1}=+1)=\epsilon_2$$ where $\epsilon_1$ and $\epsilon_2$ are respectively the probability of observing $R_i=-1$ and $R_i=0$. Similarly, if the parent gene of $g_i$ has DNA function loss, caused by mutation for instance, or its expression level is down regulated, or the protein of $g_{i-1}$ is not regulated successfully ($M_{i-1}\&R_{i-1}\&RM_{i-1}=-1$), then the downstream regulation process towards $g_i$ is likely not to be functioning. Therefore, $g_i$ would tend to be not regulated properly, namely $R_i=-1$, and hence the corresponding probability would be flipped.

	\begin{table}[!t]
		\renewcommand{\arraystretch}{1.5}
		\centering
		\caption{THE REGULATION PROCESS $e_{i-1,i}$ IN $G^*$ IS EXPRESSION}
		\label{expression_cpt}
		\begin{tabular}{cccccc}
			($RM_{i-1}\&\text{)}M_{i-1}\&R_{i-1}$ & $R_i=+1$  & $R_i=0$ & $R_i=-1$ \\
			\hline
			
			$+1$   &   $1-\epsilon_1-\epsilon_2$  &  $\epsilon_2$ &  $\epsilon_1$ \\
			$-1$   &   $\epsilon_1$ &  $\epsilon_2$   &  $1-\epsilon_1-\epsilon_2$
			
		\end{tabular}\\
		\textsuperscript{*}$0<\epsilon_1<\epsilon_2<<1-\epsilon_1-\epsilon_2$
	\end{table}
	
	\begin{table}[!t]
		\renewcommand{\arraystretch}{1.5}
		\centering
		\caption{THE REGULATION PROCESS $e_{i-1,i}$ IN $G^*$ IS REPRESSION}
		\label{repression_cpt}
		\begin{tabular}{cccccc}
			($RM_{i-1}\&\text{)}M_{i-1}\&R_{i-1}$  & $R_i=+1$  & $R_i=0$ & $R_i=-1$ \\
			\hline
			
			$+1$   & $\epsilon_1$ &  $\epsilon_2$   &  $1-\epsilon_1-\epsilon_2$  \\
			$-1$   & $1-\epsilon_1-\epsilon_2$  &  $\epsilon_2$ &  $\epsilon_1$ 
			
		\end{tabular}
	\end{table}
	
	\begin{table}[!t]
		\renewcommand{\arraystretch}{1.5}
		\centering
		\caption{THE REGULATION PROCESS $e_{i-1,i}$ IN $G^*$ IS FUNCTIONAL ACTIVATION}
		\label{functional_plus_cpt}
		\begin{tabular}{cccccc}
			($RM_{i-1}\&\text{)}M_{i-1}\&R_{i-1}$  & $RM_i=+1$  & $RM_i=0$ & $RM_i=-1$ \\
			\hline
			
			$+1$   &   $1-\epsilon_1-\epsilon_2$  &  $\epsilon_2$ &  $\epsilon_1$ \\
			$-1$   &   $\epsilon_1$ &  $\epsilon_2$   &  $1-\epsilon_1-\epsilon_2$
			
		\end{tabular}
	\end{table}
	
	\begin{table}[!t]
		\renewcommand{\arraystretch}{1.5}
		\centering
		\caption{THE REGULATION PROCESS $e_{i-1,i}$ IN $G^*$ IS FUNCTIONAL INHIBITION}
		\label{functional_minus_cpt}
		\begin{tabular}{cccccc}
			($RM_{i-1}\&\text{)}M_{i-1}\&R_{i-1}$  & $RM_i=+1$  & $RM_i=0$ & $RM_i=-1$ \\
			\hline
			
			$+1$   &   $\epsilon_1$ &  $\epsilon_2$ &  $1-\epsilon_1-\epsilon_2$  \\
			$-1$   &   $1-\epsilon_1-\epsilon_2$ &  $\epsilon_2$   &  $\epsilon_1$
			
		\end{tabular}
	\end{table}

	\subsection{Ranking the Route}
	\subsubsection{A Score Based on Conditional Probability}
	Given $(\vec{r},\vec{m},\vec{rm})$, a set of data observations of the random variables in Bayesian Network $G$ from a specific patient $s$, we could rank the path $G^*$ with the probability of observing $\vec{r}$,  $\vec{m}$ and $\vec{rm}$ conditioning on the Bayesian network model $G$, $P(\vec{R}=\vec{r},\vec{M}=\vec{m},\vec{RM}=\vec{rm}|G)$. The larger the probability, the more likely the pathway route is perturbed since the observation is highly consistent with the biological logic from $G^*$ encoded in $G$. One problem of using this probability as a measure, is that the probability will be higher if fewer data are observed. Thus the score displayed in (\ref{unsigned_route_score_formula}) given in \cite{koller2009probabilistic}, will be used instead, where the conditional probability is normalized by $P(\vec{R},\vec{M},\vec{RM} \text{ are consistent}|G)$.
	\begin{equation}
	\label{unsigned_route_score_formula}
	Score_s(G^*,\vec{r},\vec{m},\vec{rm})  = \frac{P(\vec{R}=\vec{r},\vec{M}=\vec{m},\vec{RM}=\vec{rm}\mid G)}
	{P(\vec{R},\vec{M},\vec{RM}\text{ are consistent}\mid G)}
	\end{equation}
	\begin{equation*}
	\label{unsigned_route_score_delim_formula}
	\begin{split}
	& \ \ \ \ \  \ \ \ P(\vec{R}=\vec{r},\vec{M}=\vec{m},\vec{RM}=\vec{rm} \mid G)\\
	& = \sum_{\vec{R}=\vec{r},\vec{M}=\vec{m}, \vec{RM}=\vec{rm}}P(\vec{R},\vec{M},\vec{RM})\\
	&= \sum_{\vec{R}=\vec{r},\vec{M}=\vec{m},\vec{RM}=\vec{rm}}
	\prod_{Pa^G(R_i)=\emptyset}P(R_i)
	\prod_{1\leq i <k_{G^*}}P(M_i)\\
	&\prod_{Pa^G(R_i)\neq \emptyset}P(R_i\mid Pa^G(R_i))
	\prod_{Pa^G(RM_i)\neq \emptyset}P(RM_i\mid Pa^G(RM_i))
	\end{split}
	\end{equation*}
	where $Pa^G(X)$ is the set of containing parent nodes of node $X$ in Bayesian Network $G$.
	$P(\vec{R},\vec{M},\vec{RM} \text{ are consistent}|G)$ is the probability that the random variables with observations are fully consistent with the biological logic encoded in the pathway route given $R_1=+1, M_1=+1$. For instance, suppose the pathway route only contains two genes and $g_1$ phosphorylates $g_2$, namely, $G^*: g1\overset{+p}\longrightarrow g2 $. Then we have (\ref{demo_r2_missing_formula}):
	\begin{equation}
	\label{demo_r2_missing_formula}
	\begin{split}
	P(\vec{R},\vec{M},\vec{RM}\text{ are consistent}&\mid G) \\
	= P(R_1=+1 &,M_1=+1,RM_2=+1\mid G)
	\end{split}
	\end{equation}
	(since $g_2$ is the last node in the route and the interaction $e_{12}$ is phosphorylation, then $R_2$ and $M_2$ are not included in the model). 
	
	A high score means that the path $G^*$ is highly likely to be perturbed based on the data we observe. A path $G^*$ could only get a high score if the observations, the changes in tumor cells for each gene, are highly consistent with pathway information contained in the Bayesian Network $G$. Inconsistency between data and the model would lower the score greatly since the conditional probability will be $\epsilon_1$ instead of $1-\epsilon_1-\epsilon_2$ during the calculation of the score. Advantages of this measure are 
	\begin{itemize}
		\item the analysis could be done across pathways, i.e. after merging pathways in a reasonable way, this measure could recognize a significantly meaningful route across different pathways. This could allow biologists, oncologist or doctors to see what biological processes are likely to be ``making trouble'' in the patient\rq s body.
		\item even though some observation values are flipped due to random errors from the genomic data (it is observed to be $-1$ when it actually $+1$), the whole path would still have a high score if the other genes have consistent observations.
	\end{itemize}
	The data here comes from one patient, $s$, indicating that the score is specifically tailored to patient $s$.
	
	The perturbed route could have two possible statuses, enhanced or suppressed, as we have defined in \cite{zhao2016deep}. Recall that we define a pathway route $G^*$ to be enhanced if the last gene's expression value is observed to be the same as the expected value. The expected value is calculated based on biological logic in $G^*$ by supposing $R_1=+1, M_1=+1$. The route is defined to be suppressed if the observation of the last node is opposite to the expectation. The score is easily extended to include this information, resulting in a new signed score, $sScore$, as shown in (\ref{signed_score_formula}).
	\begin{equation}
	\label{signed_score_formula}
	sScore_s(G^*,\vec{r},\vec{m})=\tilde{I}(o_{|G^*|},\dot{o}_{|G^*|})\cdot Score_s(G^*,\vec{r},\vec{m},\vec{rm})
	\end{equation}
	where $o_{G^*}$ is the observation of the last gene in the route $G^*$. If $e_{|G^*|-1,|G^*|}$ is an Expression Interaction, then $o_{|G^*|}=r_{|G^*|}$. Otherwise, $o_{|G^*|}=rm_{|G^*|}$ since $e_{|G^*|-1,|G^*|}$ is Functional interaction. $\dot{r}_{\mid G^*\mid}$ is the expected observation of the last gene in the route conditioning on $R_1=+1, M_1=+1$. Similarly, If $e_{|G^*|-1,|G^*|}$ is expression interaction, then $\dot{o}_{|G^*|}=r_{|G^*|}$. Otherwise, $\dot{o}_{|G^*|}=rm_{|G^*|}$ since $e_{|G^*|-1,|G^*|}$ is Functional interaction. Function $\tilde{I}:R^2\rightarrow R$ is defined in (\ref{indicator_tilde_I}). The signed score varies from $-1$ (highly suppressed)  to $+1$ (highly enhanced).
	\begin{equation}
	\label{indicator_tilde_I}
	\tilde{I}(x,y)=\begin{cases}
	+1 & x=y \\
	-1 &  x\neq y
	\end{cases}
	\end{equation}
	Finally we propose the measure for a whole pathway based on the route score. The pathway score for pathway $G_B$ based on data from a group of subjects $S$, $pScore_S(G_B)$, is displayed in (\ref{pathway_score_formula}): 
	\begin{equation}
	\begin{split}
	\label{pathway_score_formula}
	pScore_S(G_B)&=\frac{1}{|G_B|}\\
	&\sum_{G^*\in G_B}I(\frac{1}{|S|} \sum_{s\in S}Score_s(G^*)>\beta)
	\end{split}
	\end{equation}
	The above equation is formulated because of following reasons. The pathway could be partitioned to be several routes. We then simply measure the significance of this pathway, $G_B$, using the proportion of routes that have an average of all the patients\rq \ scores, calculated by (\ref{unsigned_route_score_formula}), that is larger than threshold $\beta$.
	%
	
	%
	
	\subsection{Data Integration}\label{data_integration}
	
	\begin{figure}[!t]
		\centering
		\includegraphics[trim=0cm 0.7cm 0cm 0cm,clip,width=0.5\textwidth]{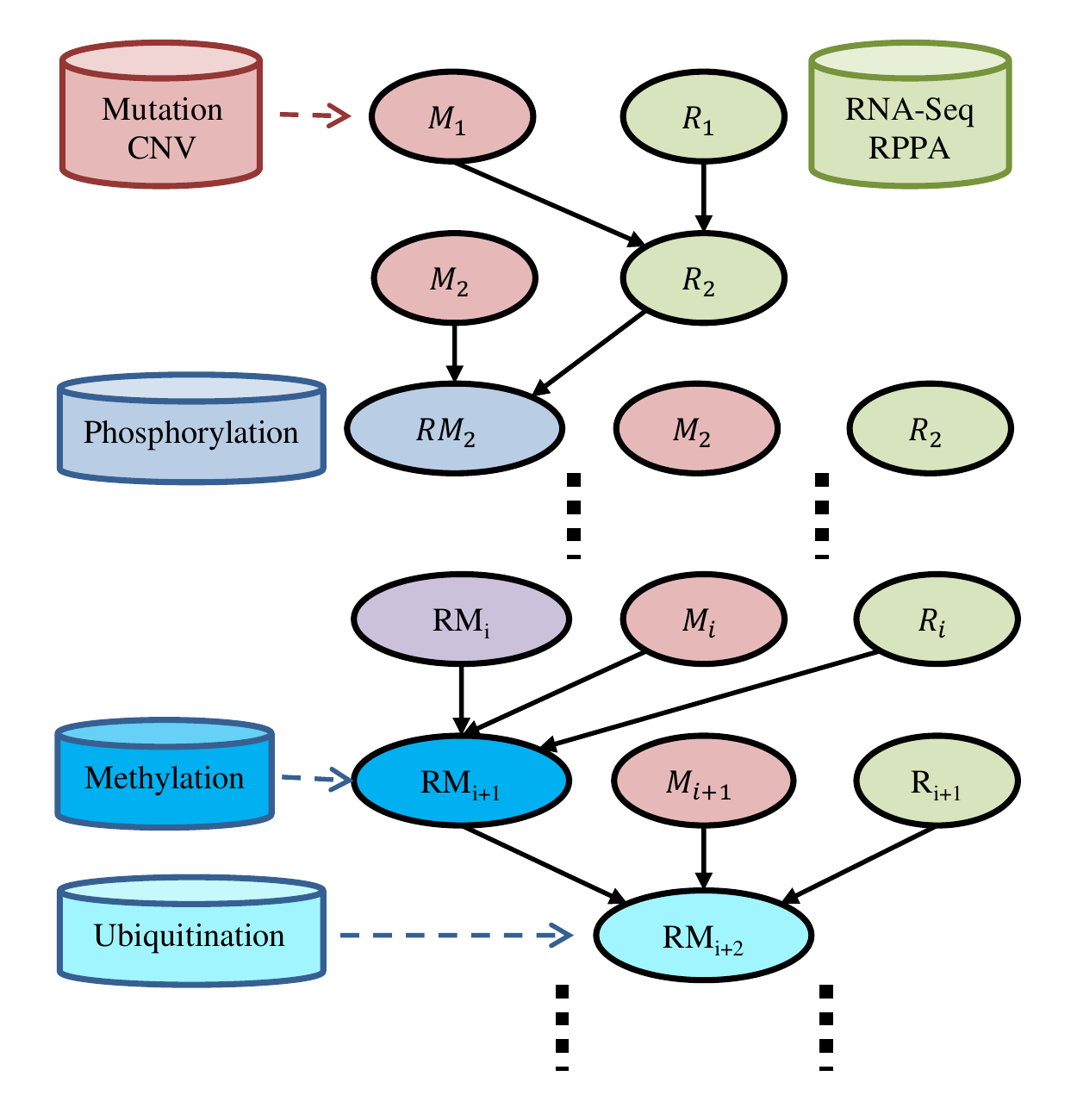}
		\caption{Data Intergration illustration. A sample Bayesian network is shown above. All the nodes with same color will have the same data resource. $RM_2$ has observation extracted from phosphorylation database because $e_{12}$ is Phosphorylation or Dephosphorylation interaction in its corresponding pathway route $G^*$. Similarly, $e_{i,i+1}$ is Methylation or Demethylation and $e_{i,i+1}$ is Ubiquitination or Deubiquitination.  }
		\label{p2_data_integration_graph}
	\end{figure}

	The observations for each variable in the Bayesian Network $G$ will come from multiple types of data, as shown in Figure~\ref{p2_data_integration_graph}. The gene expression variable $R_i$ value can be measured by many types of gene expression data, for instance, Microarray, mRNA-seq, Reverse phase protein array (RPPA) among others. Here mRNA-seq is chosen.  $R_i$\rq s observation $r_i$ is generated with log2 ratio of mRNA-seq FPKM using (\ref{rna_data_formula}). The threshold is set to be 0.5 to tolerate the random error resulting from sequence processing. If both protein data and mRNA-seq data are available for the same gene of the same patient and these two data have conflicting observation, then we use protein data observation to overwrite the one from mRNA-seq data. 
	
	$m_i$, the data observed for random variables $M_i$ is the congenital functional status for gene $g_i$.  Observation $m_i=-1$ if it can be observed from mutation or CNV data that $g_i$\rq s DNA causes function loss for the original biological process. For instance, if we observe function loss mutation from mutation database or serious copy number loss from CNV database, then $m_i=-1$. Otherwise, $m_i=+1$ standing for not observing functional deficiency in $g_i$ DNA. Mutation annotation tools will be really helpful in finding function loss mutation. 
	
	When it comes to the observation of $RM_i$, $rm_i$, the data source becomes more complex. The data source will be determined by the specific type of $e_{i-1,i}$. $RM_j$ represents different information for different interactions. The general logic is summarized by the following equation in (\ref{rm_observation_formula}). 
	\begin{equation}
	\label{rm_observation_formula}
	RM_i=Type_{i-1,i}*Tag_{i-1,i}*RawValue_i
	\end{equation}
	where $Type_{i-1,i}=+1$ if $e_{i-1,i}$ is activation (arrow) edge and $Type_{i-1,i}=-1$ if $e_{i-1,i}$ is inhibition; $Tag_{i-1,i}=+1$ if $e_{i-1,i}$ has a tag sign of ($+$), i.e. $+p,+m,+u$ or $+g$ and $Tag_{i-1,i}=-1$ if $e_{i-1,i}$ is with a tag of $-1$, i.e. $-p,-m,-u$ or $-g$. $RawValue_i=+1$ if the database shows that the gene is Phosphorylated, Methylated, Ubiquitinated or Glycosylated and $RawValue_i=-1$ if the database shows that the gene is Dephosphorylated, Demethylated, Deubiquitinated or Deglycosylated. 
	For instance, if the regulation process $e_{ij}$ is inhibition taged with phosphorylation ($+p$) and Phosphorylation data shows that $g_i$ is phosphorylated, $RM_j=-1*1*1=-1$ represents that $g_j$ is swithed off. If the data shows that the gene is not phosphorylated, $RM_i=-1*1*(-1)=+1$ indicating $g_i$ is not switched off successfully through phosphorylation. If $e_{i-1,i}$ is methylation, then the value of $rm_i$ will be determined from methylation database. The same goes with the other possible interactions: Phosphorylation, Dephosphorylation, Ubiquitination, Glycosylation etc. For the edges with no tags, we assume the edge is always working and use the fomula in (\ref{rm_obs_formula_without_tag}) instead. The formula indicates that given no function loss in $M_{i-1}$ and down expression in $R_{i-1}$, $e_{i-1,i}$ works and determines $RM_i$.
	\begin{equation}
	\label{rm_obs_formula_without_tag}
	RM_i=Type_{i-1,i}*min(M_{i-1},R_{i-1})
	\end{equation}
	

\section{Significance Analysis}
The bioinformatics field frequently uses the TCGA Breast invasive carcinoma (BRCA) data to test newly developed analysis models. We choose the same TCGA cancer data set to validate our model. Another cancer data set, Ovarian serous cystadenocarcinoma (OV) , is also analyzed with the same methodology for generality. Four types of data sets: mRNA-seq, mutation, Copy Number Variation are downloaded from https://gdac.broadinstitute.org/ for both cancer studys. Phosphorylation data are respectively extracted from recent work \cite{mertins2016proteogenomics} and \cite{zhang2016integrated}. 

The mRNA-seq data is processed as follows to obtain $r_i$, the observation for $R_i$. The cancer vs. normal paired ratios of RPKM are converted to the expression observation with (\ref{rna_data_formula}). The value for each item is mapped to a node in pathway by official gene symbol. 
\begin{equation}
\label{rna_data_formula}
r_i=\begin{cases}
+1 & log_2(\frac{TumorRPKM_i}{NormalRPKM_i})>0.5 \\
-1 & log_2(\frac{TumorRPKM_i}{NormalRPKM_i})<-0.5 \\
0 & otherwise
\end{cases}
\end{equation}

The mutation information is extracted from mutation accessor study \cite{https://doi.org/10.7908/C1F18Z2Z} and \cite{https://doi.org/10.7908/C1TT4QCX}. The mutation with a 'medium' or 'high' impact factor is encoded as function loss mutation. Other mutation observations are encoded as no function loss mutation in the data. The value for each item is mapped to a node in pathway by ncbi-protein id. Copy Number Variation (CNV) data is imported from GISTIC2 study \cite{https://doi.org/10.7908/C1NP23RQ} and \cite{https://doi.org/10.7908/C1P84B9Q}, where the copy number variation is quantified by integers varying from $-2$ to $+2$ and negative values are considered as copy number loss. CNV information determines the observation for $M$ node, $m$, along with mutation information as we discussed in section \ref{data_integration}. The value for each item is mapped to a node in pathway by official gene symbol. 


In the end, phosphorylation data is processed. $rm_i \text{(phosphorylation)}=+1$ if the same patient\rq s phosphosite iTRAQ log2 ratio is positive for $g_i$ and otherwise $rm_i(phosphorylation)=-1$. Missing values are encoded as $0$. The value for each item is mapped to a node in pathway by ncbi-protein id. However, one challenge one may encounter is that values for different residues within the same protein may be inconsistent, and KEGG pathway fails to provide sufficient information on the specific residue involved for each phosphorylation. As a result, only the consistent signal is considered in the experiment.   

All KEGG Homo Sapien pathways are used in this study, and the implementation is done mainly with R package ``KEGGgraph'' \cite{zhang2009kegggraph} and ``gRain'' \cite{H2012gRain}.


Next we do a significance analysis similar to that of PARADIGM \cite{vaske2010inference}. We will produce decoy pathways by permuting the genes in the pathway while keeping the interactions. We generate one decoy pathway for each of 308 KEGG pathways. For each pathway, we extract all possible routes in it. Then for each route, we calculate the score for each pathway by (\ref{pathway_score_formula}). We go on to rank the significant real pathways and their corresponding decoy pathway. A threshold is set to do prediction, i.e. the cases with a score higher than the threshold is predicted to be a real pathway.  After obtaining False positive rate and True positive rate with various thresholds, the resulting ROC (Receiver operating characteristic) curve can be seen in Figure~\ref{roc_plot} and Figure~\ref{ov_roc_plot}. The AUC gets to 0.77(0.72) when taking the threshold $\beta=0.52(1.00)$ for BRCA(OV). Furthermore, while taking the same value of $\beta$, the top significant pathways for Breast cancer (BRCA) and Ovarian Cancer (OV) are filtered out from the 44 signaling Homo Sapien pathways and listed in Table~\ref{brca_sig_pathway_table} and Table~\ref{ov_sig_pathway_table}. The second column corresponds to the score for the real pathway by (\ref{pathway_score_formula}). The third column contains the score for the corresponding decoy pathway. The pathways with a score less than its corresponding decoy pathway are filtered out.

\begin{figure}[!t]
	\centering
	\includegraphics[trim=0cm 1.7cm 0cm 1.5cm,clip,width=0.5\textwidth]{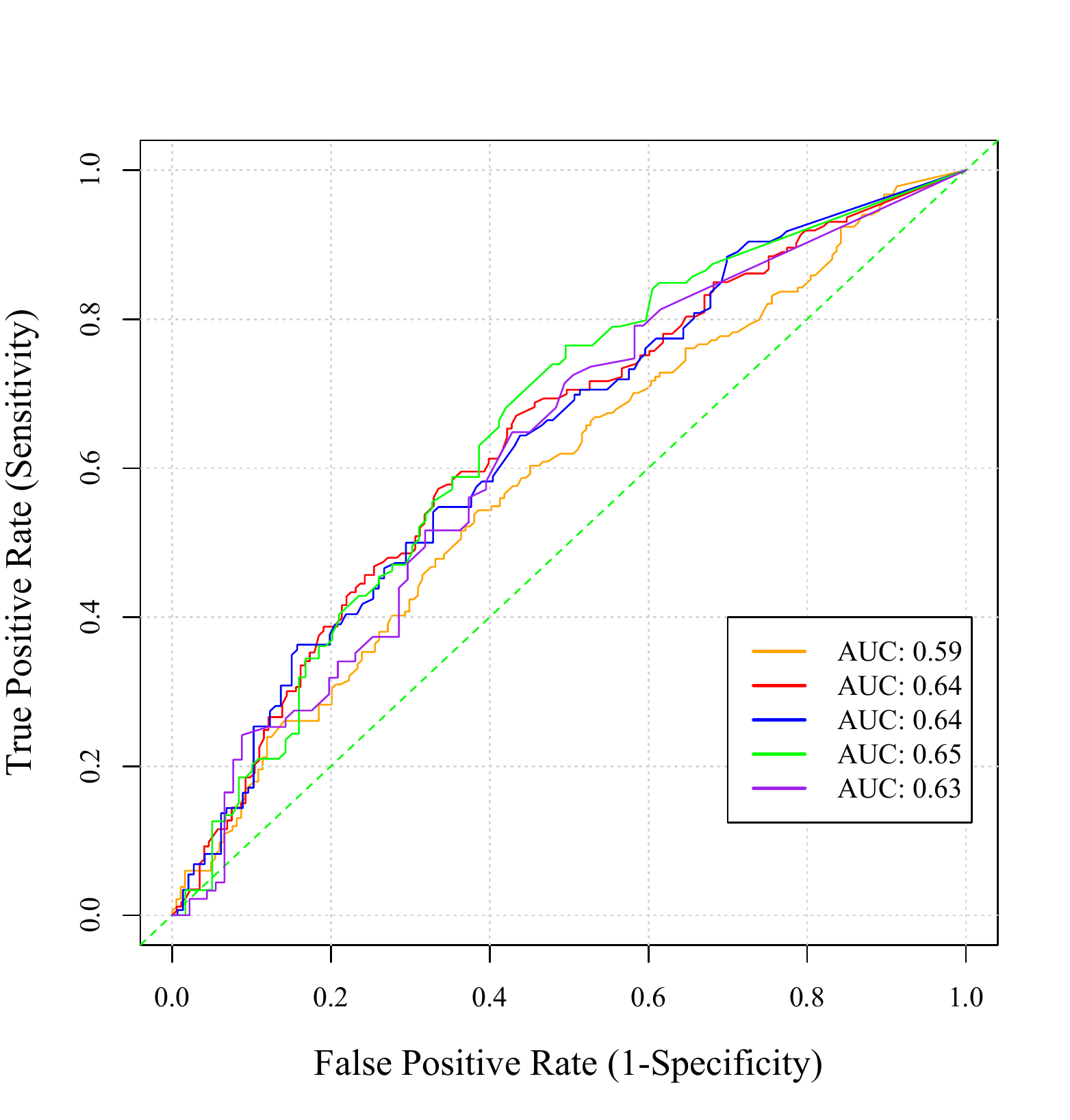}
	\caption{ROC curve for BRCA Significance study. The orange, red, blue, green and purple curve corresponds to $\beta=0.1,0.2,0.3,0.4,0.5$ seperately. The threshold is picked from [0,1] with a step of 1/10000.}
	\label{roc_plot}
\end{figure}

\begin{figure}[!t]
	\centering
	\includegraphics[trim=0cm 1.7cm 0cm 1.5cm,clip,width=0.5\textwidth]{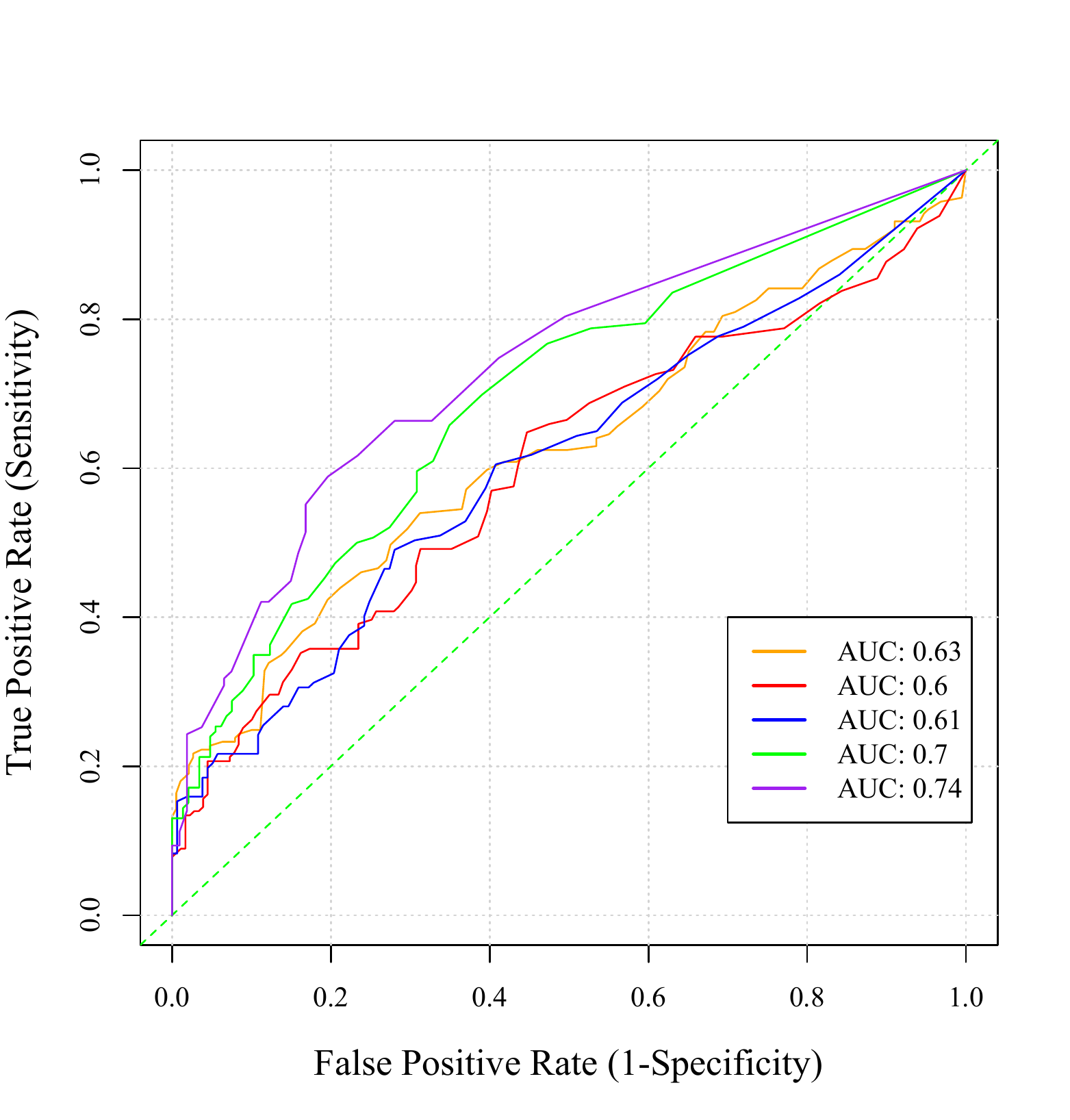}
	\caption{ROC curve for OV Significance study. The orange, red, blue, green and  dark green curve corresponds to $\beta=0.2,0.4,0.6,0.8,1.0$ seperately. The threshold is picked from [0,1] with a step of 1/10000.}
	\label{ov_roc_plot}
\end{figure}

\subsection{BRCA Pathways Verification}
The result for Breast cancer is firstly verified by biomedical literatures. The pathways in Table.\ref{brca_sig_pathway_table} is reviewed one by one. In breast cancer Neurotrophins and their receptors significantly impact tumor cell growth and metastasis through various signaling pathways according to \cite{hondermarck2012neurotrophins}. 
Vascular endothelial growth factor (VEGF) is the most prominent among the angiogenic cytokines and is believed to play a central role in the process of neovascularization, both in cancer as well as other inflammatory diseases \cite{kieran2012vegf}. 
In primary breast cancer, AMPK activity is known diminished in an estimated 90\% of cases \cite{li2015targeting}.
Dysfunction of Hippo pathway components is linked with breast cancer stem cell regulation and the connection between the disease and genetic variations in the pathway is reported in \cite{zhang2016genetic}. 
Activation of the phosphoinositide 3 kinase (PI3K)/Akt/mammalian target of rapamycin (mTOR) pathway is commonly reported in breast cancer \cite{lee2015pi3k}.
Involvement of the cAMP/protein kinase A pathway and of mitogen-activated protein kinase in the anti-proliferative effects of anandamide in human breast cancer cells is discussed in \cite{melck1999involvement}. 
\cite{buck2006tgf} discuss the key role of TGF-$\beta$ signaling. 
ErbB signaling pathway is well known frequently dysregulated in breast cancer \cite{yarden2001untangling}. 
The critical role for NF-$\kappa$B signaling pathway is discussed in \cite{shostak2011nf}. 
According to Wang and Li, LOX-1 is up-regulated by TNF in endothelial cell promoting the adhesion  and trans-endothelial migration of MDA-MB-231 breast cancer cells \cite{wang2008tumor}. 
An excellent review summarizing the role of the toll-like receptor signaling pathway on breast cancer risk, disease progression, survival, and disease recurrence is given in \cite{la2013contribution}. 
Oxytocin Receptors is related to breast cancer according to \cite{reversi2005oxytocin}. 
Blockade of Wnt/$\beta$-catenin signaling suppresses breast cancer metastasis \cite{jang2015blockade}. 
When it comes to Calcium signaling pathway, specific Ca(2+) channels reportedly play important roles in the proliferation and invasiveness of breast cancer cells \cite{azimi2014calcium}. 
\cite{kasper2009hedgehog} conclude that the inhibition of Hedgehog signalling in breast tumours may interfere with the maintenance of a putative cancer stem cell compartment and the abnormal stimulation of tumour stroma. 
Oestrogen is known to trigger the sphingolipid signaling cascade in various tissues including breast cancer \cite{sukocheva2014role}. 

Overall, we found that 94\% ($17/18$) of the pathways in Table~\ref{brca_sig_pathway_table} have some published facts implicated in breast cancer, suggesting that our analysis is producing meaningful outcomes. The only one that is not confirmed by literatures is marked with $*$ and deserves further investigation. Furthermore, we could also look into the perturbed routes in each pathway reported here and it is attached in supplementary document as Table.S1 due to its large size.


\subsection{OV Pathways Verification}

The ovarian cancer result is also validated in a similar way. 
TNF secretion by ovarian cancer cells stimulated a constitutive network consisting of cytokines, chemokines, and angiogenic factors that promoted colonization of the peritoneum and neovascularization for developing tumor deposits \cite{wang2008tumor}. 
VEGF has also been implicated in the pathogenesis of ovarian cancer according to \cite{moghaddam2012significance}. 
For MAPK pathway, MEK4 suppresses metastasis based on its downregulation in prostate and ovarian cancers with a high risk of metastasis \cite{dhillon2007map}. 
\cite{luquain2003role} identify a novel role of phospholipase D in agonist-stimulated lysophosphatidic acid synthesis by ovarian cancer cells. 
Rap1A promotes ovarian cancer metastasis via activation of ERK/p38 and notch signaling \cite{lu2016rap1a}. 
\cite{hanrahan2012genomic} reported that ovarian cancer cell lines (23.5\%) had RAS/RAF pathway aberrations. The ras-signaling pathway has attracted considerable attention as a target for anticancer therapy because of its important role in carcinogenesis \cite{adjei2001blocking}. 
\cite{szkandera2013hedgehog} highlights the crucial role of Hedgehog signaling in the development and progression of ovarian cancer. 
Findings argue that the Hippo signaling pathway defines an important pathway in progression of ovarian cancer in \cite{hall2010hippo}. 
According to \cite{cheaib2015pi3k}, phosphatidylinositol 3 kinase (PI3K) pathway is frequently altered in cancer, including ovarian cancer (OC). 
Compelling evidence suggests that NF-$\kappa$B plays a critical role in ovarian cancer in \cite{white2011genomics}. 
\cite{arend2013wnt} review the Wnt/$\beta$-catenin pathway as it relates to epithelial ovarian cancer, specifically its role in chemoresistance and its potential role as a target for chemosensitization. 
Inhibition of the JAK2/STAT3 pathway in ovarian cancer results in the loss of cancer stem cell-like characteristics and a reduced tumor burden \cite{abubaker2014inhibition}. 
The mammalian target of rapamycin (mTOR) is frequently activated in epithelial ovarian cancer, and is regarded as an attractive therapeutic target for therapy in \cite{mabuchi2011targeting}. 
\cite{corney2008role} review the role of tumor suppressor p53 and the Rb pathway in EOC with particular attention to association of p53 to high grade serous carcinomas as opposed to low grade and benign tumors.

Overall, we found that 82\% ($14/17$) of the pathways in Table~\ref{ov_sig_pathway_table} have some published facts implicated in ovarian cancer, suggesting that our analysis is producing meaningful outcomes.
Furthermore, we could also look into the perturbed routes in each pathway reported here and it is attached in supplementary document as Table.S2 due to its large size. 


\begin{table}[!t]
	\centering
	\caption{SIGNIFICANT PATHWAY REPORTED FOR BRCA}
	\label{brca_sig_pathway_table}
	
	\begin{tabular}{lcc}
		
		Pathway &	Real Score & Decoy Score\\
		\hline
		Adipocytokine signaling pathway *          & 0.02        & 0           \\
		Neurotrophin signaling pathway            & 0.008281573 & 0           \\
		VEGF signaling pathway                    & 0.007518797 & 0           \\
		AMPK signaling pathway                    & 0.005676443 & 0           \\
		Hippo signaling pathway multiple species  & 0.004608295 & 0           \\
		PI3KAkt signaling pathway                 & 0.004464286 & 0           \\
		cAMP signaling pathway                    & 0.003601441 & 0           \\
		TGFbeta signaling pathway                 & 0.003039514 & 0           \\
		ErbB signaling pathway                    & 0.002814919 & 0           \\
		NF$\kappa$B signaling pathway               & 0.002477291 & 0           \\
		TNF signaling pathway                     & 0.00210084  & 0           \\
		Tolllike receptor signaling pathway       & 0.001174168 & 0           \\
		Oxytocin signaling pathway                & 0.020889488 & 0.000673854 \\
		Hippo signaling pathway                   & 0.002457002 & 0.000117    \\
		Wnt signaling pathway                     & 0.005074161 & 0.00058548  \\
		Calcium signaling pathway                 & 0.014311736 & 0.00234192  \\
		Hedgehog signaling pathway                & 0.00487013  & 0.001623377 \\
		Sphingolipid signaling pathway            & 0.007585335 & 0.003792668 \\
		\hline
	\end{tabular}
\end{table}

\begin{table}[!t]
	\centering
	\caption{SIGNIFICANT PATHWAY REPORTED FOR OV}
	\label{ov_sig_pathway_table}
	
	\begin{tabular}{lcc}
		
		Pathway &	Real Score & Decoy Score\\
		\hline
		RIGIlike receptor signaling pathway  *                 & 0.008299381 & 0           \\
		TNF signaling pathway                                 & 0.007509387 & 0           \\
		VEGF signaling pathway                                & 0.006718925 & 0           \\
		MAPK signaling pathway                                & 0.003164489 & 0           \\
		Phospholipase D signaling pathway                     & 0.002879539 & 0           \\
		Rap1 signaling pathway                                & 0.001560912 & 0           \\
		Ras signaling pathway                                 & 0.001462766 & 0           \\
		Hedgehog signaling pathway                            & 0.000725338 & 0           \\
		Hippo signaling pathway                               & 0.000487915 & 0           \\
		PI3KAkt signaling pathway                             & 0.000434485 & 0           \\
		NF$\kappa$B signaling pathway                           & 0.000368958 & 0           \\
		Wnt signaling pathway                                 & 0.001947448 & 5.85E-05    \\
		JakSTAT signaling pathway                             & 2.27E-05    & 0           \\
		NODlike receptor signaling pathway *                   & 0.002733265 & 0.000359842 \\
		mTOR signaling pathway                                & 0.00377488  & 0.000691244 \\
		p53 signaling pathway                                 & 0.001706679 & 0.000458365 \\
		AGERAGE signaling pathway *                             & 0.006871944 & 0.001952085\\
		\hline
	\end{tabular}
\end{table}

\section{Discussion}
We further extend the existing deep pathway analysis approach by introducing more detailed information in the pathway. Unlike existing methods, the model has the ability to handle multiple types of data including  CNV, proteomics and methylation data. We first demonstrated the performance of the model through significance study with real data, and compare the result against PARADIGM and SPIA. Significant pathways reported can be verifed by current literature. In the end, we carried out a pathway route analysis (deep pathway analysis) combined with verification from biological literature. Our Bayesian based approach can be further augmented with additional statistical and machine learning methods, for example, for enhanced model selection, hypothesis test, parameter estimation

\bibliographystyle{natbib}
\bibliography{p2Bib}

\end{document}